\documentclass[letter]{spie} 

\usepackage[T1]{fontenc}
\usepackage[english]{babel}

\usepackage{graphicx} 

\usepackage{ifluatex}
\ifluatex 
    \usepackage[utf8]{luainputenc}
    \usepackage{epstopdf}
\else
    \usepackage[utf8]{inputenc}
\fi

\usepackage{amsmath,amsfonts,amssymb}
\usepackage{array}
\usepackage{mathtools}
\usepackage{textcomp}
\usepackage{xcolor}
\usepackage[super]{nth}

\graphicspath{{figures/}} 
\usepackage[colorlinks=true, allcolors=blue]{hyperref}

\title{%
	Spectral interferometric wavefront sensing: a solution for petalometry at Subaru/SCExAO
}

\author[a$\star$]{V.~Deo}
\author[a,b]{S.~Vievard}
\author[c]{M.~Lallement}
\author[d]{M.~Lucas}
\author[c]{E.~Huby}
\author[a,e]{K.~Ahn}
\author[a,b,f,g]{O.~Guyon}
\author[a]{J.~Lozi}
\author[c]{H.-D.~Kenchington Goldsmith}
\author[c]{S.~Lacour}
\author[h]{G.~Martin}
\author[i,j,k]{B.~Norris}
\author[c]{G.~Perrin}
\author[l]{G.~Singh}
\author[i,j,k]{P.~Tuthill}

\affil[a]{\small Subaru Telescope, National Astronomical Observatory of Japan, 650 N. Aohoku Pl., Hilo, HI, USA}
\affil[b]{Astrobiology Center of NINS, 2 Chome-21-1, Osawa, Mitaka, Tokyo, Japan}
\affil[c]{LESIA, Observatoire de Paris, Univ. PSL, CNRS, Sorbonne Univ., Univ. de Paris, 92195 Meudon, France}
\affil[d]{Institute for Astronomy, University of Hawai'i, 640 N. Aohoku Pl., Hilo, HI 96720, USA}
\affil[e]{Korea Astronomy and Space Science Institute, 776 Daedeok-daero, Yuseong-gu, Daejeon 34055, Republic of Korea}
\affil[f]{Steward Observatory, University of Arizona, Tucson, AZ, USA}
\affil[g]{College of Optical Sciences, University of Arizona, Tucson, AZ, USA}
\affil[h]{University Grenoble Alpes, CNRS, IPAG, 38000 Grenoble, France}
\affil[i]{Sydney Institute for Astronomy, School of Physics, The University of Sydney, NSW 2006, Australia}
\affil[j]{Sydney Astrophotonic Instrumentation Laboratories, Physics Road, University of Sydney, NSW 2006, Australia}
\affil[k]{AAO-USyd, School of Physics, University of Sydney 2006}
\affil[l]{Gemini Observatory Northern Operations Center, 670 N. A'ohoku Place Hilo, HI 96720, USA}

\authorinfo{$\star$ Corresponding author. Email: \texttt{vdeo} at \texttt{naoj} dot \texttt{org}.}

\begin{document}
\maketitle

\begin{abstract}
	The petaling effect, induced by pupil fragmentation from the telescope spider, drastically affects the performance of high contrast instruments by inducing core splitting on the PSF.
	Differential piston/tip/tilt aberrations within each optically separated fragment of the pupil are poorly measured by commonly used Adaptive Optics (AO) systems.
	We here pursue a design of dedicated low-order wavefront sensor -- or petalometers -- to complement the main AO.
	Interferometric devices sense differential aberrations between fragments with optimal sensitivity; their weakness though is their limitation to wrapped phase measurements.
	We show that by combining multiple spectral channels, we increase the capture range for petaling aberrations beyond several microns, enough to disambiguate one-wave wrapping errors made by the main AO system.
	We propose here to implement a petalometer from the multi-wavelength imaging mode of the VAMPIRES visible-light instrument, deployed on SCExAO at the Subaru Telescope.
	The interferometric measurements obtained in four spectral channels through a 7 hole non-redundant mask allow us to efficiently reconstruct differential piston between pupil petals.
\end{abstract}


			\section{INTRODUCTION}
			\label{sec:intro}

Low-wind effect\cite{Milli2018LWE,Vievard2019LWEonSCExAO,BertrouCantou2021Confusion} (LWE) has been a recurring issue for high contrast imaging on large telescopes.
The poor flushing of cold air pockets nearby the radiatively cooling secondary support spiders cause a sharp discontinuity in the wavefront between the clear apertures of the telescope pupils, often called petals.
General purpose wavefront sensors (WFSs), which perform an effective reconstruction from the gradient or the curvature of the wavefront, do poorly to reconstruct discontinuous and optically disjointed wavefronts maps.
The regularization of the sharp discontinuity induced by LWE through the adaptive optics (AO) control loop results in the spurious residual of independently evolving piston components within each separate petal -- the petaling effect (PE) -- which is not naturally occuring in Kolmogorov turbulence.
These petal modes are either mishappenly ignored by the AO controller, or sometimes worsened through a process we call petal-locking, where a one-wavelength optical path difference (OPD) step is applied in the wavefront map by the corrector\cite{BertrouCantou2021Confusion}.

Wavefront petaling dramatically hinders high-contrast imaging performance; with the PSF core effectively splitting into multiple lobes, efficient coronagraphy is compromised, and speckle structure is made unstable.
Injection into single mode waveguides for spectroscopy, interferometry, or into astrophotonic devices, is made near-impossible while strong LWE sequences are happening.

We herein present continuation of previous work\cite{Deo2021SAWFS,Deo2022SAWFSdualband} in developing an effective auxiliary WFSing scheme specifically dedicated to petal pistons, which leverages sideband visible-light photons to stabilize the PSF core in the main science channels.
This work is performed on the Subaru Coronagraphic eXtreme Adaptive Optics (SCExAO\cite{Lozi2018SCExAO,Lozi2020SCExAOStatus}) instrument at the 8.2~m Subaru Telescope, a high-contrast platform operating ~50 nights/year, a significant portion of which is allocated to science using the near-IR (1.12-2.4~\textmu m) integral field unit CHARIS\cite{Groff2017CHARISFirstLight,Kuzuhara2022Acceleratingstar}.
Our interferometric petal sensor design is implemented as a part of the VAMPIRES\cite{Norris2016VAMPIRES,Lucas2024VAMPIRES_SPIE} visible light polarimeter, one of the science modules of SCExAO, which can be used for technical photons concurrently with near-IR science.
VAMPIRES, as the rest of SCExAO's science modules, receives a twice AO-corrected light beam, being located after Subaru's facility AO3k\cite{Lozi2024AO3K} and SCExAO's AO system, which respectively provide correction onto a 3228 and a 2040-actuator deformable mirror.

We use VAMPIRES' existing sparse aperture masking (SAM) interferometric capability, a technique where the resulting interferometric PSF (e.g. fig.~\ref{fig:sam_sensing_concept}, middle panel) is the overlay of fringes between pairs of holes in a nonredundant mask, thus encoding optical path differences between the sampled sections of the pupil plane.
With this measurement performed simultaneously in 4 different narrowband filters, we are able to reconstruct the petaling wavefront error with multiple microns of capture range, a requirement to overcome spurious 1- or 2-wave wavefront petal steps induced by the Pyramid wavefront sensor (PyWFS) of SCExAO, which has a sensing band of 800-950~nm.

\section{SPARSE APERTURE WAVEFRONT SENSING}
\label{sec:exp_setup}

\begin{figure}
	\centering
	\includegraphics[width=.9\textwidth]{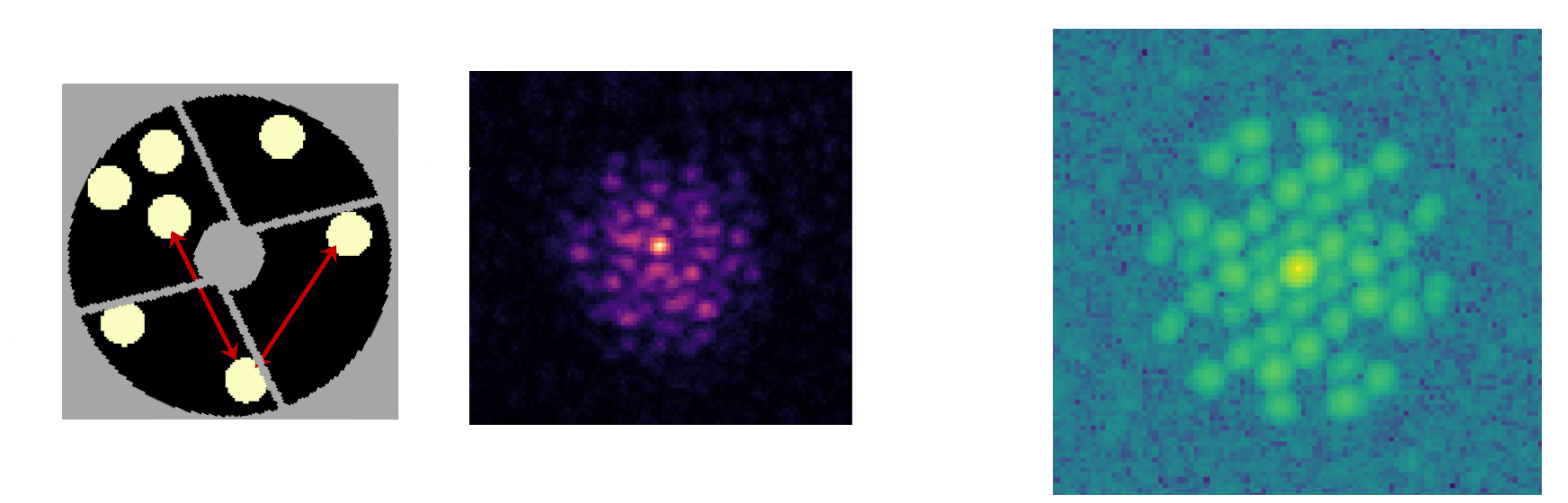}
	\caption{%
		Left: 7-hole non-redundant pupil mask, overlaid with the Subaru Telescope pupil; two of the 21 baselines are highlighted.
		Middle: typical interferogram obtained with this 7-hole mask on VAMPIRES, here at 775~nm wavelength.
		Right: complex magnitude of the Fourier transform of the interferogram, showing the 42 cross-correlation lobes corresponding to the 21 baselines of the mask.
		Figure reproduced from previous work\cite{Deo2022SAWFSdualband}.
	}
	\label{fig:sam_sensing_concept}
\end{figure}

Sparse aperture masking (SAM) interferomery\cite{Tuthill2006SparseApAO,Tuthill2013NRMInterf} has long been used to circumvent the effect of seeing. Controlling the speckle noise withing the holes of the pupil mask (fig~\ref{fig:sam_sensing_concept}, left) is essential to the success of the technique, which allows retrieving the structure of a resolved astrophysical object as one would with a multi-telescope interferometer. Originally, this was achieved with holes of the order of the atmospheric $r_0$; nowadays, coupling SAM with AO systems allows for much larger sparse holes to be used, thus dramatically increasing throughput.

Usually, self-calibrating and wavefront-agnostic observables\cite{Monnier2007PhasesInInterferometry} are used to obtain scientific measurements with nonredundant SAM interferometry; however, wavefront can be otherwise retrieved, since each baseline formed by a pair of holes creates a set of optical fringes, which phase encodes the OPD between the two holes.
We can thus easily turn a sparse aperture imager into a wavefront sensor -- a SAWFS. Note that the rationale is identical to that of fringe tracking interferometry between the piston modes of multi-telescope interferometers.

The stacked fringes of all the baselines result in an interferometric PSF, a typical one shown on fig.~\ref{fig:sam_sensing_concept} (center).
Computing the Fourier transform of the PSF, we obtain a correlogram such as on fig.~\ref{fig:sam_sensing_concept} (right): a collection of disjoint lobes describing the autocorrelation of the electric field in the entrance pupil, i.e. the system's (instantaneous) optical transfer function (OTF).
The non-redundancy ensures that each OTF lobe corresponds to exactly one pair of holes from the SAM.
The complex phase sampled at the peak of each lobe is a direct function of the OPD over the wavefront baseline corresponding to this lobe.
Put in an equation, the correlogram value at the peak of the lobe corresponding to holes $i$ and $j$ ($1 \leq i < j \leq N_\text{holes}$) interfering together is:
\begin{equation}
	\label{eq:1}
	C_{ij} \propto \exp\left(i(\psi_{ij} + \phi_i - \phi_j)\right)
\end{equation}
From which, after discarding the complex amplitude, and calibrating the bias term $\psi_{ij}$, we simply obtain the differential phase:
\begin{equation}
	\label{eq:2}
	\phi_{ij} = \phi_i - \phi_j \text{ mod } 2\pi.
\end{equation}

With 7 holes, the measurement of the 21 $\phi_{ij}$ ought to be of rank 6, corresponding to the wavefront on all 7 holes minus overall piston; by applying a further transformation, we may reconstruct up to 6 wavefront modes of choice. This is done very classically, by feeding the 21 $\phi_{ij}$ measurements --the preprocessed SAWFS output-- into a classical least squares AO controller after subtracting the offset induced by instrumental aberration and possible scientific biases from an extended astrophysical source.

This WFS design has several drawbacks, mainly using photons that could otherwise be used for science, and limited throughput from the use of the sparse mask itself -- approximately 15\% with the 7 hole mask.
However, it comes with other benefits as compared to other low-order WFSing techniques. In theory, the linearity of the measurement is perfect, affected only by speckle noise within the holes and chromatic smearing of the correlogram lobes with the spectral bandwidth; and the noise propagation is optimal over the photons that effectively reach the detector, with interferometric techniques achieving the theoretical photon noise propagation bound of 1 rad$^2$ of variance per mode per photon received\cite{Guyon2005Limits}.


One key limitation of interferometric fringe phase readout is the limitation of dynamic range by phase wrapping, as explicited on eqs.~\ref{eq:1} and~\ref{eq:2}. For a central measurement wavelength $\lambda$, the unambiguous capture range, in units of OPD, is $\left[-\dfrac{\lambda}{2}, +\dfrac{\lambda}{2}\right]$. While using visible light, this amounts only to $\sim\pm$350-400~nm, an insufficient quantity to measure possible 1-wave petal lockups at the wavelength of the AO system. This is where we leverage two-wavelength interferometry\cite{Martinache2022SAM,Houairi2009TwoWLEstimator}, a well known technique in fringe tracking and long baseline interferometry.
Given two measurements $\phi(\lambda_1)$ and $\phi(\lambda_2)$ of an achromatic OPD-induced phase at two different wavelengths $\lambda_1$ and $\lambda_2 (> \lambda_1)$, we can retrieve an unwrapped measurement at the equivalent pseudo-wavelength
\begin{equation}
	\label{eq:3}
	\Lambda = \dfrac{\lambda_1\lambda_2}{\lambda_2 - \lambda_1},
\end{equation}
with the retrieved effective phase being
\begin{equation}
	\label{eq:4}
	\phi(\Lambda) = \phi(\lambda_2) - \phi(\lambda_1)\text{ mod } 2 \pi,
\end{equation}
with an effective unambiguous OPD capture range now extended to $\left[-\dfrac{\Lambda}{2}, +\dfrac{\Lambda}{2}\right]$, albeit at the cost of a comparably increased noise propagation (in OPD), since we would now use:
\begin{equation}
	\delta_\Lambda = \dfrac{\Lambda}{2\pi} \phi(\Lambda)
\end{equation}
rather than the OPD computed from the original phase measurements:
\begin{equation}
	\delta_{\lambda_1, \lambda_2} = \dfrac{1}{2\pi} \dfrac{\lambda_1\phi(\lambda_1) + \lambda_2\phi(\lambda_2)}{2}
\end{equation}

\section{Experimental setup}

			\subsection{VAMPIRES}
			\label{sec:experimental:vampires}

\begin{figure}
	\centering
	\includegraphics[width=0.8\textwidth]{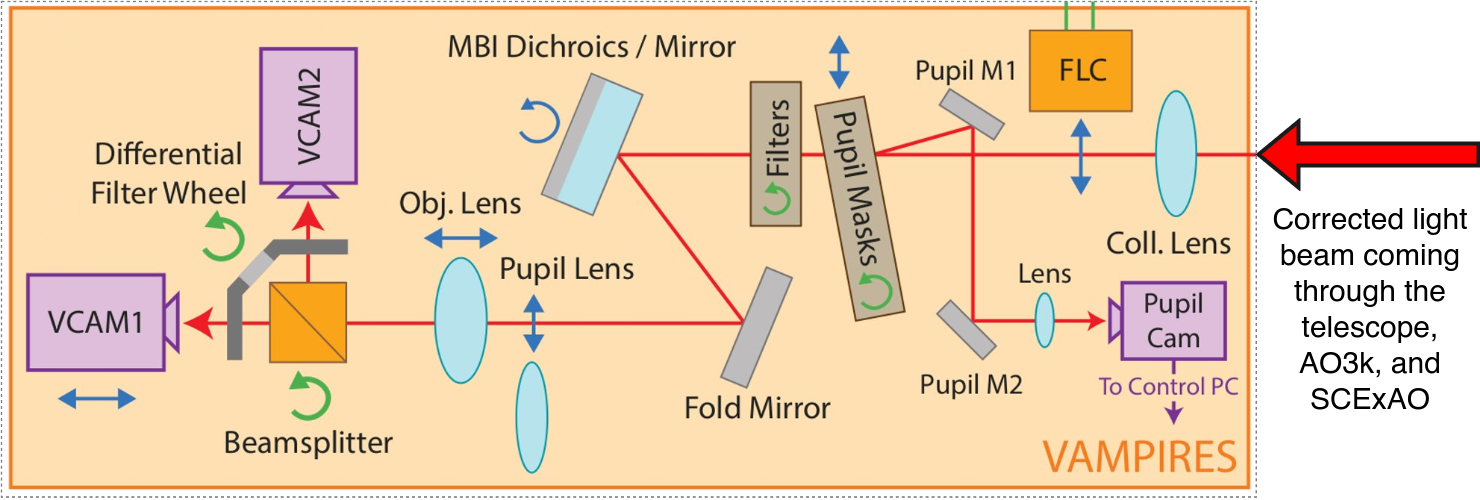}
	\caption{%
	Layout of the VAMPIRES module on SCExAO.
	The beam (600-800~nm) is received from the right side, after being corrected from atmospheric turbulence by the AO3k and SCExAO deformable mirrors.
	The 7 hole NRM is held within the \textbf{Pupil Masks} wheel; the beam then reflects off the MBI dichroic stack (fig.~\ref{fig:mbi_diagram}).
	For SAWFS, only one detector is needed and the polarizing beamsplitter may be moved out entirely.
	Figure reproduced from Lucas \emph{et al}\cite{Lucas2024VAMPIRES_SPIE}.
	}
	\label{fig:vampires_diagram}
\end{figure}

VAMPIRES (Visible Aperture Masking Polarimetric Imager for Resolved Exoplanetary Structures) is the main visible-light instrument available for scientific observation on SCExAO.
Its general optical layout is shown on fig.~\ref{fig:vampires_diagram}.
As part of its scientific capabilities, VAMPIRES includes non-redundant pupil masks with 7, 9 and 18 holes\cite{Norris2016VAMPIRES,Uyama2020VampiresHAlpha}. It receives the visible light between the AO188 and SCExAO WFSs cutoffs, from 550 to 800~nm.
Other than non-redundant masking, scientific capabilities include a suite of coronagraphs\cite{Doelman2024VAMPIRESVortex,Lucas2022VAMPIRESCoronagraph}, triple-differentiation polarimetric imaging, H$\alpha$ and SII spectral differential imaging, with a 3'' field-of-view at a resolution of 17 to 20 mas, using two simultaneous cameras imaging both polarizations and/or continuum and narrowband lines.

\begin{figure}
	\centering
	\includegraphics[width=0.39\textwidth]{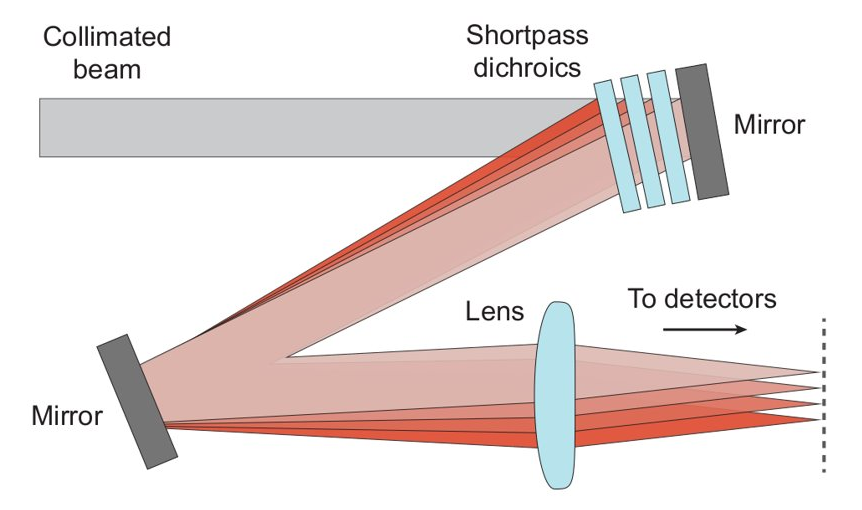}%
	\includegraphics[width=0.37\textwidth]{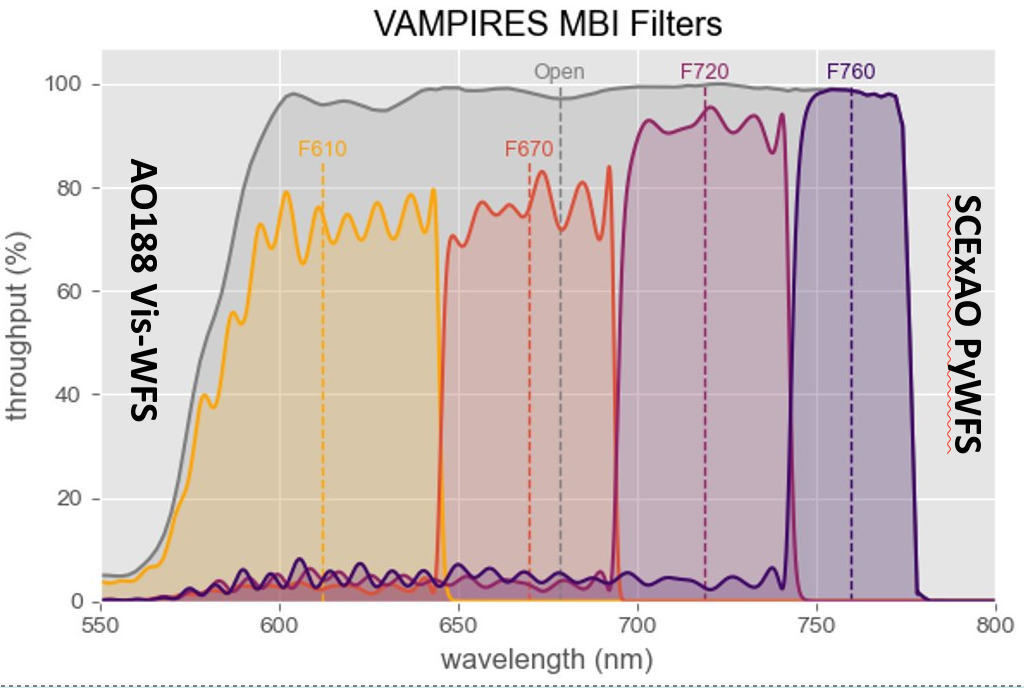}%
	\includegraphics[width=0.24\textwidth]{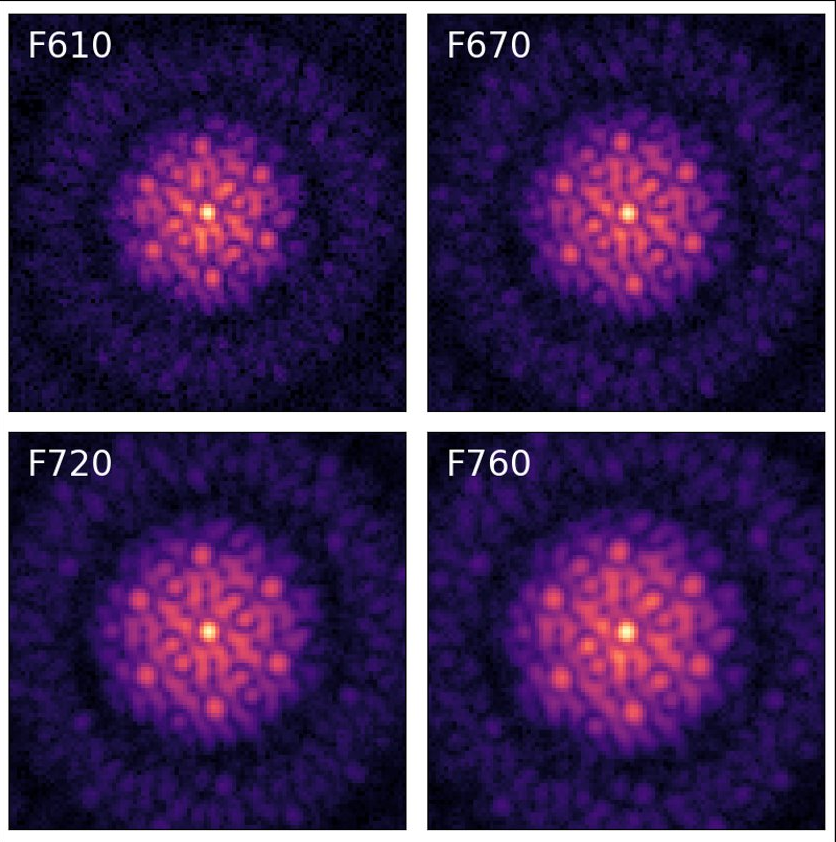}%
	\caption{%
	Details on the MBI dichroic stack design (left). Spectral characterization of the resulting effective bandpass filters F610, F670, F720 and F760 (center). Sample 7-hole mask interferograms acquired from the SCExAO calibration source (right); the 4 crops are taken from the same detector readout, cropped from a much larger 2228x1104~px total detector readout area.
	Left and center panels reproduced from Lucas \emph{et al}\cite{Lucas2024VAMPIRES_SPIE}.
	}
	\label{fig:mbi_diagram}
\end{figure}

In 2023, a detector upgrade was conducted, from 512$\times$512 EMCCD cameras to photon-counting qCMOS cameras with 4K detectors and significantly increased readout speeds.
This upgrade, its design, and its technical and scientific capabilities, is detailed in Lucas et al\cite{Lucas2024VAMPIRES_SPIE}.
As part of this REVampiNG\footnote{%
	\emph{Readout Enhanced VAMPIres; Now Great!.}
}, a so-called dichroic stack system was included, as shown on fig.~\ref{fig:mbi_diagram}, which comprises of a tube with a mirror at the back end, upon which are stacked shortpass dichroics angled to offset the beam onto different portions of the detectors.
Using 3 low angle-of-incidence dichroics, we achieve 4 effective spectral channels (fig.~\ref{fig:mbi_diagram}, center) imaged simultaneously on the detectors, which we call the multiband imaging mode (MBI).

MBI allows us to leverage two-wavelength interferometry as described in eq.~\ref{eq:3} and~\ref{eq:4} and will be the mode selected for SAWFS.
It is otherwise also compatible with all pupil masks, coronagraphic, and polarimetric features of VAMPIRES.

\subsection{A four-colored SAWFS}
\label{sec:reduction:single}

By combining the MBI optics with the one of the sparse aperture masks (only the 7-hole mask is used in the results presented herein), we effectively obtain the capability to run a SAWFS at four neighboring wavelengths simultaneously.
We refer the reader to previous work\cite{Deo2021SAWFS,Deo2022SAWFSdualband} for more details on how we calibrate and preprocess the data from each spectral channel, essentially taking the Fourier transform of the interferogram, removing static biases, and calibrating the location of the correlation lobes shown on fig.~\ref{fig:sam_sensing_concept}.

The precise location of the Fourier plane lobes in each spectral channel against the plate scale of the instrument yields the actual (weighted) central wavelengths of the channels, for the spectrum of the calibration source of SCExAO: we measure $\lambda_{1-4} = 628, 674, 725$ and 770~nm. In particular, the shift of $\lambda_1$ from 610~nm (nominal) to 628~nm (calibrated) is well explained by the sharp drop in illumination of the supercontinuum laser as it reaches its short-wavelength limit.

Using two wavelength interferometry recombination, as described in eqs.~\ref{eq:3} and~\ref{eq:4} and combining spectral channels with all six pairwise combinations, we obtain the 6 pseudo-wavelengths listed in table~\ref{tab:pseudo_wl}.
\begin{table}
	\centering
	\caption{Table: pseudo-wavelengths resulting from pairwise combinations the VAMPIRES MBI filters.}
	\label{tab:pseudo_wl}
	\begin{tabular}{c|ccc}
	& $\lambda_2 = 674$~nm & $\lambda_3 = 725$~nm & $\lambda_4 = 770$~nm \\
	\hline
	$\lambda_1 = 628$~nm & $\Lambda_{12} = 9116$~nm & $\Lambda_{13} = 4659$~nm & $\Lambda_{14} = 3580$~nm \\
	$\lambda_2 = 674$~nm &  & $\Lambda_{23} = 9529$~nm & $\Lambda_{24} = 5895$~nm \\
	$\lambda_3 = 725$~nm &  &  & $\Lambda_{34} = 15460$~nm
	\end{tabular}
\end{table}
Of note, the shortest pseudo-wavelength is only 3.6~\textmu m, resulting in a maximum range of $\pm 1.8$~\textmu m before running into phase-wrapping issues. This is generally sufficient for the SCExAO PyWFS, where a 1-wave petal error would cause $\sim$1~\textmu m wavefront discontinuities.
$\pm 1.8$~\textmu m could however be somewhat limiting in the case of petal locking induced by an infrared WFS, such as the pyramid wavefront sensor deployed in Subaru's facility AO3k.

By taking the median of the 6 OPD measurements at the 6 pseudo-wavelengths, we manage to increase the overall effective wavelength to the third smallest pseudo-wavelength, $\Lambda_{24} = 5.9~$\textmu m, i.e. to having a capture range of $\pm 2.95$~\textmu m wavefront discontinuities -- with a minimal noise penalty compared to simply taking the average of the 6 reconstructed OPDs and being limited to $\pm 1.8$~\textmu m. \textbf{The overall noise propagation is a standard deviation increase of a factor 13.8 on the photon noise error budget}, as compared to a similar SAWFS with a single spectral channel and a $\pm \lambda / 2$ capture range.
We quantitatively noted that this median solution with 4 channels had marginally better noise propagation than discarding the measurements at $\lambda_1$ entirely, which is otherwise theoretically sound.

As to summarize, the process of reconstructing petaling from quad-band interferograms is the following:
\begin{itemize}
	\setlength\itemsep{-0.3em}
	\item Retrieve the phase difference $\phi_{ijk}$ for each baseline $ij$ and each wavelength $k$.
	\item Recombine the spectral channels using two-wavelength interferometry and median selection, to obtain baseline OPDs $\delta_{ij}$.
	\item Reconstruct the wavefront map value onto the 7 holes, $\delta_i, 1\leq i \leq 7$.
	\item Reconstruct up to 6 wavefront modes from the holes' OPD value, in the case of this paper  we focus on the 4 value of petal pistons.
\end{itemize}

\subsection{Control scheme for closed loop tests}
\label{sec:experimental:control}

\begin{figure}[h!]
\centering
\includegraphics[width=.98\textwidth]{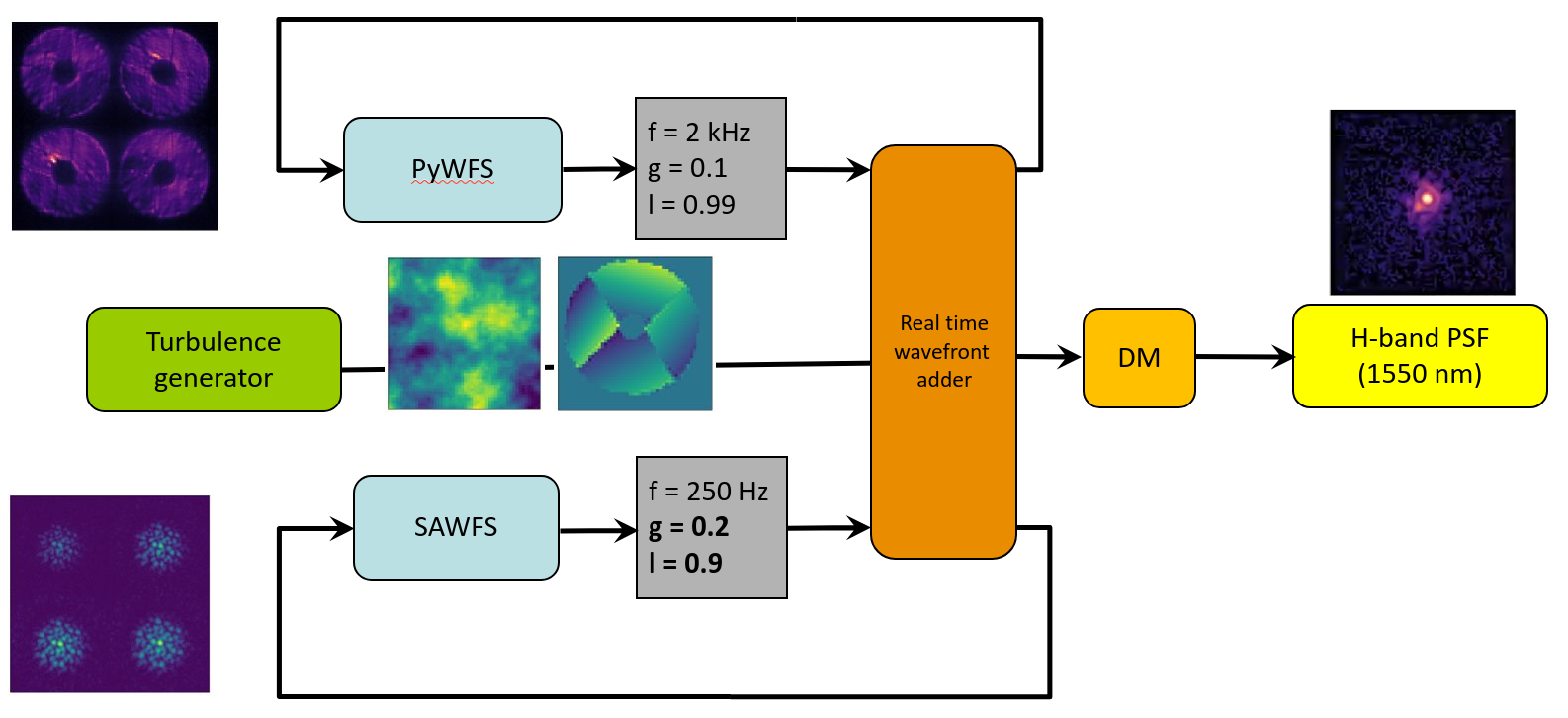}
\caption{Diagrammatic flow of closed-loop experiments. The turbulence simulator, PyWFS, and SAWFS are run in parallel without particular synchronization, with their outputs summed and sent to the DM in real-time for each update.}
\label{fig:cl_diagram}
\end{figure}

We test the SAWFS in a closed-loop setup against the visible light PyWFS of SCExAO and the internal turbulence simulator (rendered of the 2040-actuator MEMS DM).
These tests are performed running both the PyWFS and SAWFS loops simultaneously yet asynchronously, and summing the correction onto SCExAO's common path mirror, as shown on fig.~\ref{fig:cl_diagram}.

The PyWFS loop is run with typical parameters as used in testbed mode or on-sky -- with a leaky integrator with 0.1 gain ($g$) and 0.99 leakage ($l$), where we use the common autogressive equation:
\begin{equation}
\mathbf{c}[k] = g \times \mathbf{\delta c}[k] + l \times \mathbf{c}[k-1],
\end{equation}
where $\mathbf{c}$ is the command vector and $\mathbf{\delta c}$ the increment resulting from the closed loop measurements.
The PyWFS loop corrects $\sim$900~wavefront modes; the modulation radius is 3~$\lambda$/D.
We purposefully avoid implementing any advanced sensor fusion scheme -- and instead set the SAWFS loop to a high-gain high-leak control setting, with a gain of 0.2 and leakage of 0.9.
Our design here relies on having the SAWFS send ``petal kicks'', resetting a heavily petaled wavefront to flatness, and then leak quickly so that the weakly sensistive PyWFS loop absorbs the correction sent by the SAWFS, thus transfering the appropriate correction from the SAWFS integrator output into the PyWFS integrator output.
Having the PyWFS run 8x faster than the SAWFS favors the success of this loop-to-loop transfer effect.

We use two pertubatory datasets: one von Kármán dataset, which amounts to 800~nm RMS turbulence with typical peak-to-valley of 3.0~\textmu m; and a random walk dataset of pure LWE (over all 11 piston-tip-tilt modes), defined with 4~kHz resolution and amounting statistically to 50~nm RMS of LWE evolution each millisecond (or 1.6~\textmu m each second).
These two datasets are hereafter referred to as the VK and LWE turbulences.
For either of these, we acquired several closed loop sequences: open-loop/turbulence only, PyWFS loop closed, and 4 instances of PyWFS and SAWFS loops closed with varying illumination on the SAWFS, corresponding to effective on-sky R-band magnitudes of 2.0, 4.5, 7.0, and 9.2 (given the VAMPIRES zero-point over all four MBI filters of 1.8~$10^{10} e^-/s$ (see \footnote{\url{www.naoj.org/Projects/SCEXAO}}, \cite{Lucas2024VAMPIRES_SPIE}) -- and with a 7H NRM throughput of 15\%). It should be noted that while the simulated illumination is reduced on the SAWFS, it is not on the PyWFS, which is still effectively seeing a bright source with negligible noise effect for all the closed loop timeseries measured with this dual-loop setup.

\section{Experimental results}
\label{sec:results}


	\subsection{Linearity tests}
	\label{sec:lin_ramps}

\begin{figure}
	\centering
	\includegraphics[width=0.99\textwidth]{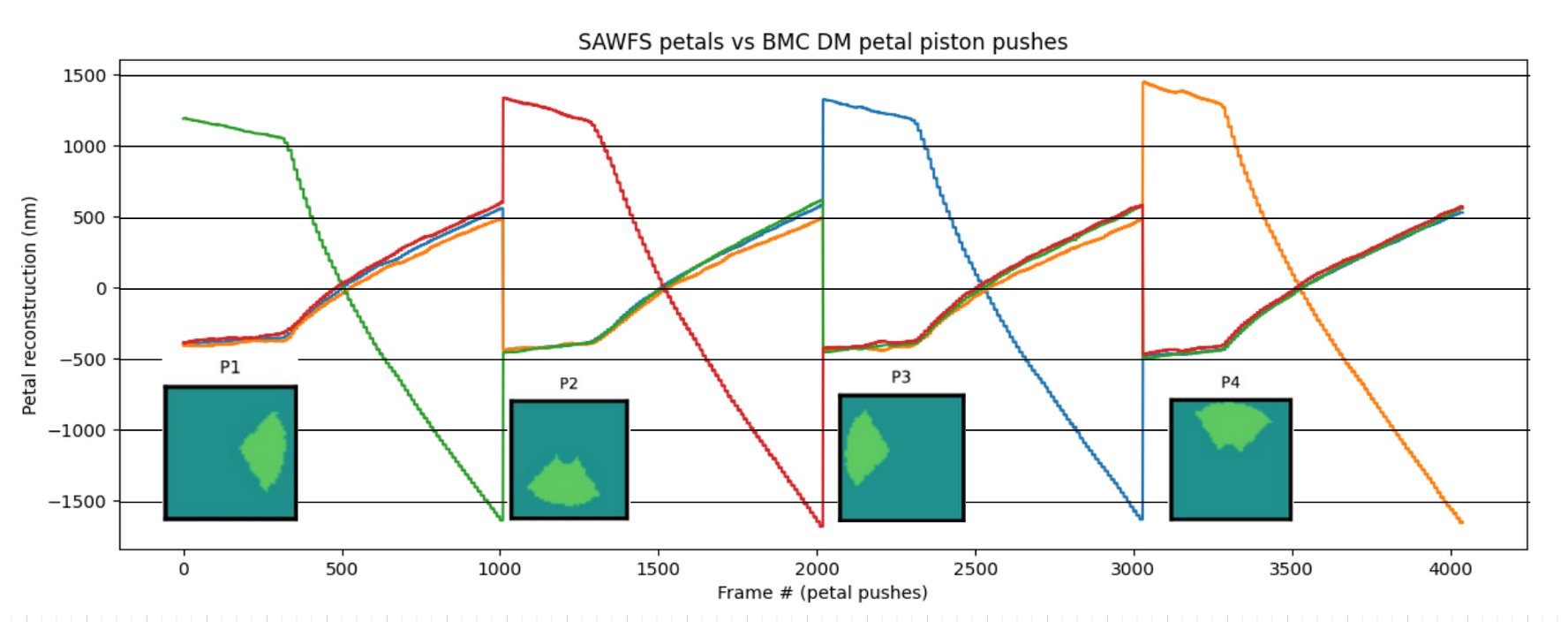}
	\caption{SAWFS reconstructed measurements when pushing each of the four petal piston modes on the SCExAO DM. From left to right across four sequences, corresponding to the mode pushed (shown in inserts within the square circumscribing the DM), the given petal piston mode traverses a push-pull ramp from \textbf{-2~\textmu m to +2~\textmu m of OPD.}
	Piston subtraction in the DM controller, and zero-mean reconstruction, results in a split of the introduced push value of +75\% of the push on the controlled mode, and -25\% on the 3 other petal piston modes.}
	\label{fig:petal_ramps}
\end{figure}

We perform linearity and dynamic range testing of our 4-wavelength SAWFS against a large range of petal piston pushes on all 4 petals of the Subaru telescope pupil. Results are shown on fig.~\ref{fig:petal_ramps}.
The experimental sequence is the following: from -2~\textmu m to +2~\textmu m of OPD, we send a ramp of petal piston values to each petal mode and measure the reconstruction through the quad-wavelength SAWFS. Due to global piston removal, the values applied to the DM and measured reflect 75\% of the value of the push sent on the mode driven, and -25\% of the same value on the 3 other petal piston modes.

As it turns out, this experiment is more of a DM characterization than a characterization of the SAWFS itself.
Saturation of large negative (pull) values is caused by the DM controller reaching its saturation level of 0V (as this is a unipolar MEMS DM). The saturation effect in the left third of each ramp of fig.~\ref{fig:petal_ramps} is progressive instead of perfectly flat, as the additional (non-flat) offset map of the DM causes a progressive saturation of the wavefront as the petal pull increases.
Over the unsaturated part of the OPD ramp, we observe a power law curve in the response ramps, which can be pinned to MEMS DM displacement not quite being a quadratic function of voltage. Altogether, the effective dynamic range of the MEMS DM as configured during those ramps is not capable of inducing phase wrap error at all at the interferometric pseudo-wavelength of 5.9~\textmu m, and confirms excellent measurement stability and dynamic range of the design.

From the saturation of the DM at low DC voltage, up to the maximum push we have exerted during this experiment, each petal mode driven spans at least 2.5~\textmu m displacement without notable artifacts, instabilities, or wraparounds. Adding in the zero-average effect, this amounts to a range of tested discontinuities spanning at least 3.3~\textmu m in total (approx from -1.3 to +2.0 given the asymmetry of the DM range), and confirming our capture range is greater than this amount.

	\subsection{Closed loop tests}
	\label{sec:res_cl}

	\begin{figure}
		\centering
		\includegraphics[width=0.8\textwidth]{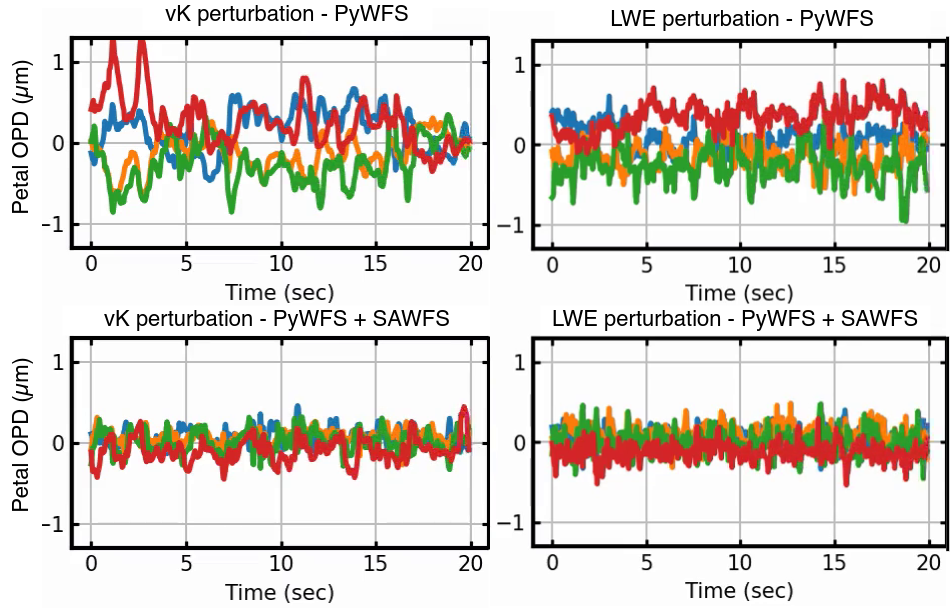}
		\caption{Timeseries of petal piston mode values during 20 second close loop sequences, on the VK and LWE perturbation datasets described in sect.~\ref{sec:experimental:control}. Each line drawn corresponds to one of the petal piston modes, their values adding to 0. 
		Top row: aberration and correction with the 2~kHz PyWFS loop; bottom row: correction with the 2~kHz PyWFS and the 250~Hz SAWFS.}
		\label{fig:res_petal_curves}
	\end{figure}

We show on fig.~\ref{fig:res_petal_curves} the results of the closed loop experiments described in sec.~\ref{sec:experimental:control} above, running only the PyWFS, or jointly with the SAWFS, against the VK and LWE turbulence sequences.
These results are shown as timeseries of the \emph{true} petal piston values, i.e. as measured on the DM total aberration map.
This measurement is enabled in the lab, by the fact that the disturbing aberration, VK or LWE, is injected on the DM and therefore known. Wavefront statistics are sampled at 2~kHz, the PyWFS loop frequency.

The top panels of fig.~\ref{fig:res_petal_curves} show the petal timeseries while only closing the PyWFS loop. The loop parameters described in sec.~\ref{sec:experimental:control} and the turbulence parameters were picked so that petal locking would be induced by the PyWFS controller consistently.
While this is not what we try to achieve while operating SCExAO on sky, it is a reasonable choice to try and replicate on-sky conditions when significant LWE episodes are in progress.
For the VK perturbation, a 1-wave gap is created by the PyWFS splitting the pupil into 2 vs 2 petals (red-blue vs orange-green) for more than half of the 20 second sequence.
Remarkably, we have two short bursts at the beginning of the sequence where the turbulence causes the pyramid to cause 2-wave excursions between petals as a preferential operating point!
For the LWE sequence, the PyWFS consistently stabilizes a petal lock of 1-wave. Since petal are natively present in the introduced aberration, there is less opportunity for the pyramid to correct the LWE properly, since petal steps are not natively included in the PyWFS control eigenmodes.

Now, on the bottom panels of fig.~\ref{fig:res_petal_curves}, we show identical sequences, now with the SAWFS running in parallel to the PyWFS, as described in sec.~\ref{sec:experimental:control} and shown on fig.~\ref{fig:cl_diagram}. The petal excursions are significantly reducted, even though a few sub-second bursts are visible in the VK dataset; for the LWE dataset, we see a few transients < 0.5 seconds with a peak petal gap of $\sim$0.5~\textmu m. In both case, the rejection of petaling at low frequencies is visibly improved, at the cost of high frequency fluctuations, induced by the high noise propagated during the multi-wavelength recombination process.

\begin{figure}[h!]
	\centering
	\includegraphics[width=.8\textwidth]{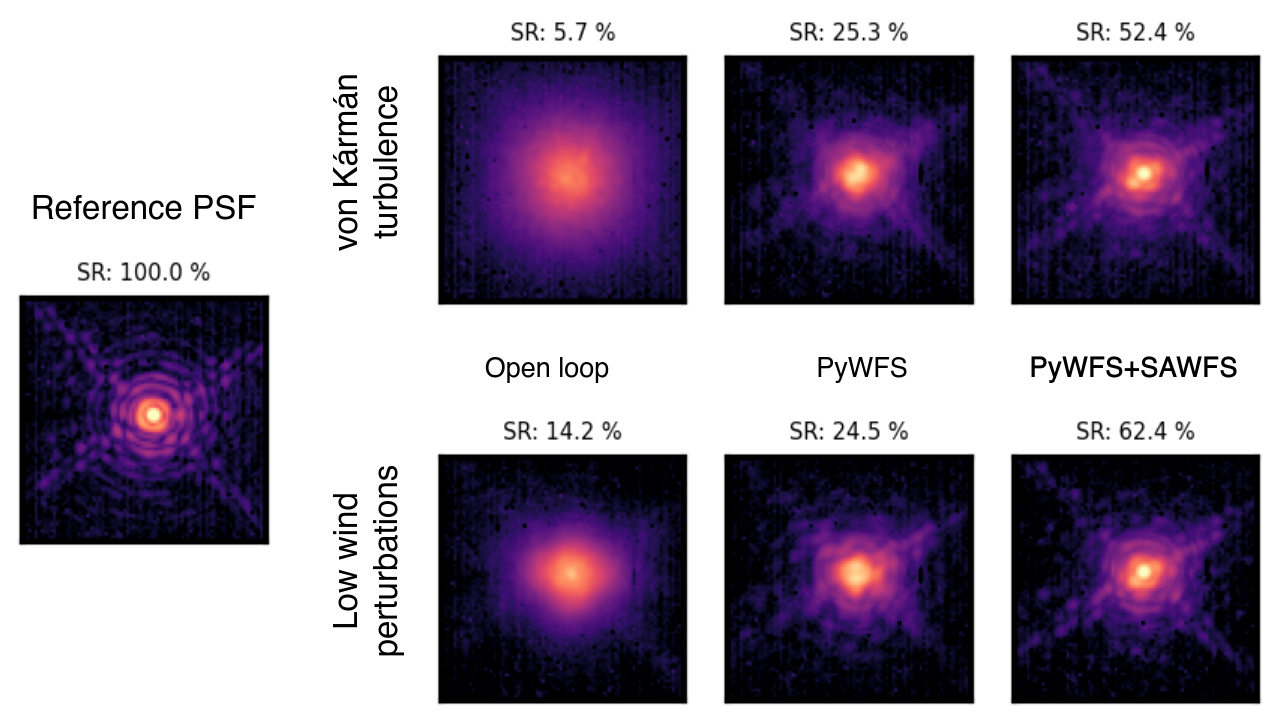}
	\caption{Long exposures H-band PSFs during the closed-loop experiment batch.}
	\label{fig:cl_psfs}
\end{figure}

Quantitative results for the runs shown on fig.~\ref{fig:res_petal_curves} are reported in tables~\ref{tab:results_vk} and~\ref{tab:results_lwe}. The corresponding long exposure PSFs for the 20 second sequences are shown on fig.~\ref{fig:cl_psfs}.
\textbf{For the VK dataset, the SAWFS improves petal residual from 842 down to 388 nm RMS, and Strehl from 25.3 up to 52.4\%.
For the LWE dataset, the SAWFS improves petal residual from 804 down to 373 nm RMS, and Strehl from 24.5 up to 62.4\%.
}
Adding the SAWFS doesn't remove petaling entirely, and some PSF elongation and distortion of the first Airy ring is still visible in the righmost panels of fig.~\ref{fig:cl_psfs}, however with no common measure with the smeared, multi-lobe PSFs that can be observed and persist throughout strong petal-locking sequences.

	\subsection{A sprinkle of photon noise}
	\label{sec:res_noise}

To complement the closed-loop tests shown in sec.~\ref{sec:res_cl}, where the SAWFS receives an effective illumination of a R=2 star, we performed similar closed loop sequences while reducing the illumination received by the SAWFS, down to magnitude R=9.2.
Practically, we reduced the VAMPIRES' camera exposure time, rather than dimming the light source; therefore, for the results presented in tables.~\ref{tab:results_vk} and~\ref{tab:results_lwe}, the effective magnitude seen by the SAWFS increases, but not that seen by the PyWFS.

Generally speaking, the faintness of the source bears no impact in the ultimate Strehl ratio achieved in H-band (except for LWE aberration, R=9.2), although the increase in petal WFE is apparent in the numbers as the source dims.
We note that the SAWFS petal measurement is consistently smaller than the true petal WFE -- there are probably multiple factors at play in this regard (the sensor fusion scheme, and the slower SAWFS loop), but we believe the main argument is tip-tilt or speckle noise at the scale of the SAWFS holes, which causes changes in the fringe structure beyond the basic interferometric model we use to describe the problem. This should probably be analyzed much more in further studies.

\begin{figure}[h]
	\centering
	\includegraphics[width=0.25\textwidth]{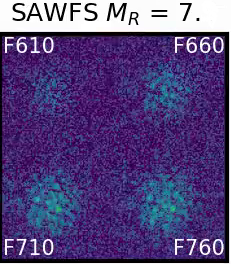}\hspace*{2em}
	\includegraphics[width=0.25\textwidth]{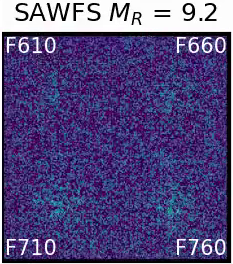}
	\caption{%
		Sample SAWFS interferograms at 4~ms exposure time for illuminations equivalent to R=7.0 and R=9.2, as used in the experiments shown in sect.~\ref{sec:res_noise}.
	}
	\label{fig:sawfs_noisy_psfs}
\end{figure}

It takes a large amount of photons to achieve good measurements with the SAWFS, in particular with the large noise propagation penalty incurred as a tradeoff to dynamic range; however, it takes fairly few photons to efficiently nudge the PyWFS into its best behavior. We show sample interferograms for the R=7.0 and 9.2 cases on figure~\ref{fig:sawfs_noisy_psfs} to provide a sense of the data, respectively with 17000 and 2500 photons in the frame.

\begin{table}[h!]
	\centering
	\caption{Closed loop sequence results (VK perturbation)}
	\vspace*{.5em}
	\label{tab:results_vk}
	\begin{tabular}{cccc}
		Dataset & True petal WFE & SAWFS petal measurement & H-band Strehl\\
		 & (nm RMS) & (nm RMS) & (\%)\\
		 \hline
		 Open loop & 866 & 749 & 5.7\\
		 PyWFS only & 842 & 651 & 25.3\\
		 \hline
		 PyWFS + SAWFS (R=2.0) & 388 & 211 & 52.4\\
		 PyWFS + SAWFS (R=4.5) & 375 & 283 & 52.7\\
		 \hline
		 PyWFS + SAWFS (R=7.0) & 388 & 288 & 54.9\\
		 PyWFS + SAWFS (R=9.2) & 472 & 343 & 54.3\\
	\end{tabular}
	\vspace*{.5em}
	\caption{Closed loop sequence results (LWE perturbation)}
	\vspace*{.5em}
	\label{tab:results_lwe}
	\begin{tabular}{cccc}
		Dataset & True petal WFE & SAWFS petal measurement & H-band Strehl\\
		 & (nm RMS) & (nm RMS) & (\%)\\
		 \hline
		 Open loop & 522 & 551 & 14.2\\
		 PyWFS only & 804 & 689 & 24.5\\
		 \hline
		 PyWFS + SAWFS (R=2.0) & 373 & 233 & 62.4\\
		 PyWFS + SAWFS (R=4.5) & 421 & 404 & 57.9\\
		 \hline
		 PyWFS + SAWFS (R=7.0) & 430 & 404 & 61.1\\
		 PyWFS + SAWFS (R=9.2) & 515 & 363 & 50.4\\
	\end{tabular}
\end{table}

			\section{CONCLUSION AND PERSPECTIVES}
			\label{sec:conclusion}

With thick and solid secondary support spiders, a design choice made with prime focus instrumentation in mind, the Subaru Telescope is the poster child of petaling issues on 8~m telescopes, and an ideal learning place to prepare diffraction limited instrumentation for extremely large telescopes.

We have shown that using non-redundant, sparse aperture masks with a limited number of large holes provides the necessary information to reconstruct just the petaling modes, and complement the main PyWFS AO system for these few improperly sensed modes.
This strategy of having a petalometer in complement to the main AO system is proving effective, pending that a sideband (or a fraction of the light in the AO band) can be spared for this use.

We demonstrate in this paper that a SAWFS using multiple neighboring spectral channels, used with the pseudo-wavelength phase retrieval technique, provides sufficient sensitivity and dynamic range to mitigate the petal-locking behavior of a PyWFS, even with the SAWFS deployed at the blue end of the available spectrum.

In the case of SCExAO, we demonstrate in the lab the restoring of petal-locking down to within 2-magnitudes of the limit of the PyWFS, while being available during all near-infrared science cases: loop settings that with the PyWFS alone yield good high order correction but a constantly petal-split core can be restored to nominal performance with the assistance of the quad-wavelength SAWFS.

For future work, we seek to redesign the NRM used toward an engineering optimized solution: a mask with only 4 holes in the 4 petals of the Subaru pupil, maximizing throughput.
We seek to perform on sky engineering in real conditions and confirm the behavior shown on the SCExAO calibration source, in hopes to considerably improve the efficiency of AO3k + SCExAO for infrared science cases, and for paving the way to ELT-sized petalometric solutions.

			\acknowledgments
			\label{sec:acknowledgements}

Based on data collected at Subaru Telescope, which is operated by the National Astronomical Observatory of Japan. 
The authors recognize the very significant cultural role and reverence that the summit of Maunakea has always had within the Hawaiian community; and acknowledge the great privilege being granted to conduct observations from this mountain, and the responsibilities attached to it.
The development of SCExAO is supported by the Japan Society for the Promotion of Science (Grant-in-Aid for Research \#23340051, \#26220704, \#23103002, \#19H00703, \#19H00695 and \#21H04998), the Subaru Telescope, the National Astronomical Observatory of Japan, the Astrobiology Center of the National Institutes of Natural Sciences, Japan, the Mt Cuba Foundation and the Heising-Simons Foundation. VD acknowledges support from NASA support grant \#80NSSC19K0336 and the Heising-Simons Foundation.

\bibliographystyle{spiebib} 
\bibliography{doctorat} 

\begin{thebibliography}{10}

\bibitem{Milli2018LWE}
Milli, J., Kasper, M., Bourget, P., Pannetier, C., Mouillet, D., Sauvage,
  J.-F., Reyes, C., Fusco, T., Cantalloube, F., Tristam, K., Wahhaj, Z.,
  Beuzit, J.-L., Girard, J.~H., Mawet, D., Telle, A., Vigan, A., and N'Diaye,
  M., ``{Low wind effect on VLT/SPHERE: impact, mitigation strategy, and
  results},'' {\em Proc. SPIE}~{\bf 10703},  752 -- 771 (2018).

\bibitem{Vievard2019LWEonSCExAO}
Vievard, S., Bos, S.~P., Cassaing, F., Ceau, A., Guyon, O., Jovanovic, N.,
  Keller, C.~U., Lozi, J., Martinache, F., Mary, D., Montmerle-Bonnefois, A.,
  Mugnier, L., N'Diaye, M., Norris, B., Sahoo, A., Sauvage, J.-F., Snik, F.,
  Wilby, M.~J., and Wong, A., ``Overview of focal plane wavefront sensors to
  correct for the {Low Wind Effect} on {Subaru/SCExAO},'' in [{\em
  6\textsuperscript{th} AO4ELT conference-Adaptive Optics for Extremely Large
  Telescopes}{\nolinebreak\hspace{0.1em}]},  (2019).

\bibitem{BertrouCantou2021Confusion}
Bertrou-Cantou, A., Gendron, {\'{E}}., Rousset, G., Deo, V., Ferreira, F., and
  Vidal, F., ``Confusion in differential piston measurement with the pyramid
  wavefront sensor,'' {\em Astronomy \& Astrophysics}~{\bf 658}(A49) (2022).

\bibitem{Deo2021SAWFS}
Deo, V., Vievard, S., Cvetojevic, N., Norris, B., Guyon, O., Lozi, J., Ahn, K.,
  Huby, E., Lacour, S., Martinache, F., Skaf, N., and Tuthill, P., ``Wavefront
  sensing using non-redundant aperture masking interferometry: tests and
  validation on {Subaru/SCExAO},'' {\em Proc. SPIE}~{\bf 11823},  11823--39
  (2021).

\bibitem{Deo2022SAWFSdualband}
Deo, V., Vievard, S., Cvetojevic, N., Ahn, K., Guyon, O., Huby, E., Lacour, S.,
  Lozi, J., Martinache, F., Norris, B., Skaf, N., and Tuthill, P.,
  ``Controlling petals using fringes: wavefront sensing through sparse aperture
  masking interferometry at {Subaru/SCExAO},'' {\em Proc. SPIE}~{\bf 12185},
  12185--34 (2022).

\bibitem{Lozi2018SCExAO}
Lozi, J., Guyon, O., Jovanovic, N., Goebel, S., Pathak, P., Skaf, N., Sahoo,
  A., Norris, B., Martinache, F., N'Diaye, M., Mazin, B., Walter, A.~B.,
  Tuthill, P., Kudo, T., Kawahara, H., Kotani, T., Ireland, M.~J., Cvetojevic,
  N., Huby, E., Lacour, S., Vievard, S., Groff, T.~D., Chilcote, J.~K., Kasdin,
  J., Knight, J., Snik, F., Doelman, D., Minowa, Y., Clergeon, C., Takato, N.,
  Namura, M., Currie, T., Takami, H., and Hayashi, M., ``{SCExAO}, an
  instrument with a dual purpose: perform cutting-edge science and develop new
  technologies,'' {\em Proc. SPIE}~{\bf 10703} (2018).

\bibitem{Lozi2020SCExAOStatus}
Lozi, J., Guyon, O., Vievard, S., Sahoo, A., Deo, V., Jovanovic, N., Norris,
  B., Martinod, M.-A., Mazin, B., Walter, A., Fruitwala, N., Steiger, S.,
  Davis, K., Tuthill, P., Kudo, T., Kawahara, H., Kotani, T., Ireland, M.,
  Anagnos, T., Schwab, C., Cvetojevic, N., Huby, E., Lacour, S., Barjot, K.,
  Groff, T.~D., Chilcote, J., Kasdin, J., Martinache, F., Laugier, R., N'Diaye,
  M., Knight, J., Males, J., Bos, S., Snik, F., Doelman, D., Miller, K.,
  Bendek, E., Belikov, R., Pluzhnik, E., Currie, T., Kuzuhara, M., Uyama, T.,
  Nishikawa, J., Murakami, N., Hashimoto, J., Minowa, Y., Clergeon, C., Ono,
  Y., Takato, N., Tamura, M., Takami, H., and Hayashi, M., ``{Status of the
  SCExAO instrument: recent technology upgrades and path to a system-level
  demonstrator for PSI},'' {\em Proc. SPIE}~{\bf 11448},  100 (2020).

\bibitem{Groff2017CHARISFirstLight}
Groff, T., Chilcote, J., Brandt, T., Kasdin, J., Galvin, M., Loomis, C., Rizzo,
  M., Knapp, G.~R., Guyon, O., Jovanovic, N., Lozi, J., Currie, T., Takato, N.,
  and Hayashi, M., ``First light of the charis high-contrast integral-field
  spectrograph,'' {\em Proc. SPIE}~{\bf 10400},  1040016 (2017).

\bibitem{Kuzuhara2022Acceleratingstar}
Kuzuhara, M., Currie, T., Takarada, T., Brandt, T.~D., Sato, B., Uyama, T.,
  Janson, M., Chilcote, J., Tobin, T., Lawson, K., Hori, Y., Guyon, O., Groff,
  T., Lozi, J., Vievard, S., Sahoo, A., Deo, V., Jovanovic, N., Ahn, K.,
  Martinache, F., Skaf, N., Akiyama, E., Norris, B.~R., Bonnefois, M.,
  Helminiak, K.~G., Kudo, T., McElwain, M.~W., Wagner, K., Wisniewski, J.,
  Knapp, G.~R., Kwon, J., Nishikawa, J., Serabyn, E., Hayashi, M., and Tamura,
  M., ``Direct imaging discovery and dynamical mass of a substellar companion
  orbiting an accelerating {Hyades} {Sun-like} star with {SCExAO/CHARIS},''
  {\em The Astrophysical Journal Letters}~{\bf In press.} (2022).

\bibitem{Norris2016VAMPIRES}
Norris, B., Schworer, G., Tuthill, P., Jovanovic, N., Guyon, O., Stewart, P.,
  and Martinache, F., ``{The VAMPIRES instrument: imaging the innermost regions
  of protoplanetary discs with polarimetric interferometry},'' {\em Monthly
  Notices of the Royal Astronomical Society}~{\bf 447},  2894--2906 (01 2015).

\bibitem{Lucas2024VAMPIRES_SPIE}
Lucas, M., Norris, B., Guyon, O., Bottom, M., Deo, V., Vievard, S., Lozi, J.,
  Ahn, K., Ashcraft, J., Currie, T., Doelman, D., Kudo, T., Lilley, L.,
  Leboulleux, L., Millar-Blanchaer, M., Safonov, B., Tuthill, P., Uyama, T.,
  and Zhang, M., ``Visible-light high-contrast imaging polarimetry at subaru,''
  {\em Proc. SPIE}  (2024).

\bibitem{Lozi2024AO3K}
Lozi, J., Ahn, K., Blue, H., Chun, A. an~Clergeon, C., Deo, V., Guyon, O.,
  Hattori, T., Minowa, Y., Nishiyama, S., Ono, Y., and Vievard, S., ``Ao3k at
  subaru: first on-sky results of the facility extreme-ao,'' {\em Proc.
  SPIE}~{\bf 13097},  13097--1 (2024).

\bibitem{Tuthill2006SparseApAO}
Tuthill, P., LLoyd, J., Ireland, M., Martinache, F., Monnier, J., Woodruff, H.,
  ten Brummelaar, T., Turner, N., and Townes, C., ``Sparse-aperture adaptive
  optics,'' {\em Proc. SPIE}~{\bf 6272},  62723A (2006).

\bibitem{Tuthill2013NRMInterf}
Tuthill, P., ``The unlikely rise of masking interferometry: leading the way
  with 19\textsuperscript{th} century technology,'' {\em Proc. SPIE}~{\bf
  8845},  884502 (2013).

\bibitem{Monnier2007PhasesInInterferometry}
Monnier, J.~D., ``Phases in interferometry,'' {\em New Astronomy Reviews}~{\bf
  51}(8-9),  604--616 (2007).

\bibitem{Guyon2005Limits}
Guyon, O., ``Limits of adaptive optics for high-contrast imaging,'' {\em
  Astrophysical Journal}~{\bf 629}(1) (2005).

\bibitem{Martinache2022SAM}
Martinache, F., Cvetojevic, N., and Deo, V., ``Fizeau-interferometry fringe
  tracking solutions for giant segmented telescope petal modes,'' {\em Proc.
  SPIE}~{\bf 12185},  12185A--1 (2022).

\bibitem{Houairi2009TwoWLEstimator}
Houairi, K. and Cassaing, F., ``Two-wavelength interferometry: extended range
  and accurate optical path difference analytical estimator,'' {\em J. Opt.
  Soc. Am. A}~{\bf 26}(12),  2503 (2009).

\bibitem{Uyama2020VampiresHAlpha}
Uyama, T., Norris, B., Jovanovic, N., Lozi, J., Tuthill, P.~G., Guyon, O.,
  Kudo, T., Hashimoto, J., Tamura, M., and Martinache, F., ``{High-contrast
  H$\alpha$ imaging with Subaru/SCExAO + VAMPIRES},'' {\em Journal of
  Astronomical Telescopes, Instruments, and Systems}~{\bf 6}(4),  1--17 (2020).

\bibitem{Doelman2024VAMPIRESVortex}
Doelman, D., Lucas, M., Nishie, Y., Hirai, T., Ishiguro, T., Lozi, J., Vievard,
  S., Deo, V., Snik, F., Norris, B., and Guyon, O., ``A broadband vector vortex
  coronagraph for {SCExAO/VAMPIRES},'' {\em Proc. SPIE}~{\bf 13100},  13100--99
  (2024).

\bibitem{Lucas2022VAMPIRESCoronagraph}
Lucas, M., Bottom, M., Guyon, O., Lozi, J., Norris, B., Deo, V., Vievard, S.,
  Ahn, K., Skaf, N., and Tuthill, P., ``A visible-light coronagraph for
  {SCExAO/VAMPIRES},'' {\em Proc. SPIE}~{\bf 12184},  12184--163 (2022).

\end{thebibliography}

\appendix
\renewcommand{\theequation}{\thesection.\arabic{equation}}
\numberwithin{equation}{section}
\renewcommand{\thefigure}{\thesection.\arabic{equation}}
\numberwithin{figure}{section}

\end{document}